\documentclass[fleqn,twoside]{article}
\usepackage{amsfonts,amssymb,cite}

\newcommand{\bls}[1]{\renewcommand{\baselinestretch}{#1}}

\def\noi{\noindent}

\makeatletter
\renewcommand{\section}{\@startsection{section}{1}{0pt}%
        {-3.5ex plus -1ex minus -.2ex}{2.3ex plus .2ex}%
        {\large\bf\protect\raggedright}}

\renewcommand{\subsection}{\@startsection{subsection}{2}{0pt}%
        {-3ex plus -1ex minus -.2ex}{1.4ex plus .2ex}%
        {\normalsize\bf\protect\raggedright}}

\renewcommand{\thesubsubsection}%
        {\arabic{section}.\arabic{subsection}.\arabic{subsubsection}.}

\renewcommand{\@oddhead}{\raisebox{0pt}[\headheight][0pt]{%
   \vbox{\hbox to\textwidth{\rightmark \hfil \rm \thepage \strut}\hrule}}}
\renewcommand{\@evenhead}{\raisebox{0pt}[\headheight][0pt]{%
   \vbox{\hbox to\textwidth{\thepage \hfil \leftmark \strut}\hrule}}}
\newcommand{\heads}[2]{\markboth{\protect\small\it #1}{\protect\small\it #2}}

\makeatother

\def\prepno#1#2
    {\thispagestyle{empty}
    \noi    \unitlength=1mm
    	\begin{picture}(178,10)
            \put(177,15){\llap{\bf #1}}
            \put(177,10){\llap{\small\rm #2}}
        \end{picture}
          }             

\newcommand{\Title}[1]{\noi {\uppercase{\Large #1}} \\}

\def\Aunames#1{\noi{\large\bf #1}}
\def\auth#1{${}^{#1}$}
\def\Addresses#1{\medskip\noi \protect
	\begin{description}\itemsep -3pt
        {\it #1} \end{description}}
\def\addr#1#2{\item[${}^{#1}$]{\it #2}}

\newcommand{\Abstract}[1]{\vskip 2mm \begin{center}
        \parbox{16.4cm}{\small\noi #1} \end{center}\medskip}
\newcommand{\PACS}[1]{\begin{center}{\small PACS: #1}\end{center}}

\newcommand{\email}[2]{\footnotetext[#1]{e-mail: #2}
		\addtocounter{footnote}{1}}

\newcommand{\Ref}[1]{Ref.\,\cite{#1}}

\def\nq{\hspace*{-1em}}
\def\nqq{\hspace*{-2em}}
\def\nhq{\hspace*{-0.5em}}

\def\inch{\hspace*{1in}}

\def\Jl#1#2{{\it #1\/} {\bf #2},\ }

\def\ApJ#1 {\Jl{Astroph. J.}{#1}}
\def\CQG#1 {\Jl{Class. Quantum Grav.}{#1}}
\def\DAN#1 {\Jl{Dokl. AN SSSR}{#1}}
\def\GC#1 {\Jl{Grav. \& Cosmol.}{#1}}
\def\GRG#1 {\Jl{Gen. Rel. Grav.}{#1}}
\def\JETF#1 {\Jl{Zh. Eksp. Teor. Fiz.}{#1}}
\def\JETP#1 {\Jl{Sov. Phys. JETP}{#1}}
\def\JHEP#1 {\Jl{JHEP}{#1}}
\def\JMP#1 {\Jl{J. Math. Phys.}{#1}}
\def\NPB#1 {\Jl{Nucl. Phys.}{B\ #1}}
\def\NP#1 {\Jl{Nucl. Phys.}{#1}}
\def\PLA#1 {\Jl{Phys. Lett.}{#1A}}
\def\PLB#1 {\Jl{Phys. Lett.}{#1B}}
\def\PRD#1 {\Jl{Phys. Rev.}{D\ #1}}
\def\PRL#1 {\Jl{Phys. Rev. Lett.}{#1}}

\def\al{&\nhq}
\def\lal{&&\nqq {}}
\def\eq{Eq.\,}
\def\eqs{Eqs.\,}
\def\beq{\begin{equation}}
\def\eeq{\end{equation}}
\def\bear{\begin{eqnarray}}
\def\bearr{\begin{eqnarray} \lal}
\def\ear{\end{eqnarray}}
\def\earn{\nonumber \end{eqnarray}}

\def\nnn{\nonumber\\ \lal }
\def\nnnv{\nonumber\\[5pt] \lal }

\def\eql{\al =\al}

\def\tst{\textstyle}

\def\fract#1#2{{\tst\frac{#1}{#2}}}

\def\half{{\fract{1}{2}}}

\def\e{{\,\rm e}}
\def\d{\partial}

\def\sign{\mathop{\rm sign}\nolimits}
\def\diag{\mathop{\rm diag}\nolimits}

\def\const{{\rm const}}

\newcommand{\vars}[1]{\left\{\begin{array}{ll}#1\end{array}\right.}

\def\({\left(}
\def\){\right)}

\def\Ref#1{(\ref{#1})}

\def\mn{_{\mu\nu}}

\def\mN{_\mu^\nu}

\def\R{{\mathbb R}}
\def\cR{{\cal R}}

\def\wh{wormhole}
\def\whs{wormholes}

\def\bhs{black holes}

\def\ssph{static, spherically symmetric}
\def\asflat{asymptotically flat}

\heads{K.A. Bronnikov, S.V. Chervon and S.V. Sushkov}{Wormholes supported by chiral fields}

\bls{0.98}
\begin{document}
\twocolumn[
\prepno{ArXiv: yymm.nnnn}{}
\vspace*{-1cm}

\Title{Wormholes supported by chiral fields}

\Aunames {Kirill A. Bronnikov\auth{a,1}, Sergey V. Chervon\auth{b,2}, and
      Sergey V. Sushkov\auth{c,3}
      }

\Addresses{
\addr a {Center of Gravitation and Fundamental Metrology,
     VNIIMS, Ozyornaya St. 46, Moscow 117361, Russia; \\
     Institute of Gravitation and Cosmology,
         PFUR, Miklukho-Maklaya St. 6, Moscow 117198, Russia  }
\addr b {Department of Theoretical Physics, Ulyanovsk State University,
        Leo Tolstoy St. 42, 432000 Ulyanovsk, Russia;
        Department of General Physics, Ulyanovsk State Pedagogical
        University, \\
        Lenin's 100 years Sq., 4, 432700 Ulyanovsk, Russia}
\addr c
{Department of General Relativity and Gravitation, Kazan State University,\\
    Kremlyovskaya St. 18, Kazan 420008, Russia;\\
 Department of Mathematics, Tatar State University of Humanities and
    Education,\\ Tatarstan St. 2, Kazan 420021, Russia}
      }

\Abstract
 {We consider static, spherically symmetric solutions of general relativity
  with a nonlinear sigma model (NSM) as a source, i.e., a set of scalar
  fields $\Phi = (\Phi^1,...,\Phi^n)$ (so-called chiral fields) parametrizing
  a target space with a metric $h_{ab}(\Phi)$. For NSM with zero potential
  $V(\Phi)$, it is shown that the space-time geometry is the same as with a
  single scalar field but depends on $h_{ab}$. If the matrix $h_{ab}$ is
  positive-definite, we obtain the Fisher metric, originally found for a
  canonical scalar field with positive kinetic energy; otherwise we obtain
  metrics corresponding to a phantom scalar field, including singular and
  nonsingular horizons (of infinite area) and wormholes. In particular,
  the Schwarzschild metric can correspond to a nontrivial chiral field
  configuration, which in this case has zero stress-energy. Some explicit
  examples of chiral field configurations are considered. Some qualitative
  properties of NSM configurations with nonzero potentials are pointed out.
  }

] 
\email 1 {kb20@yandex.ru}
\email 2 {sv\_chervon@rambler.ru}
\email 3 {sergey\_sushkov@mail.ru; sergey.sushkov@ksu.ru}

\section{Introduction}

  For many years scalar fields have been an object of great interest for
  at least two reasons. The first one is quite pragmatic: models with scalar
  fields are relatively simple, and therefore it appeared possible to
  study them in detail and then extrapolate the results to more realistic and
  complicated models. Another reason has a more physical basis. Though so far
  there in no direct observational evidence, it is generally supposed that
  there exist fundamental scalar fields of great importance for the structure
  of the Universe. As a bright example, one may mention numerous inflationary
  models in which inflation in the early Universe is typically driven by a
  fundamental scalar field called an inflaton.

  Studies of scalar fields in general relativity trace back to the paper
  by Fisher \cite{Fis} who first found a static spherically symmetric
  solution of the Einstein-scalar equations with a single scalar field
  $\phi$. Then this solution was repeatedly rediscovered (see a historical
  review in \cite{ch97m}) and discussed from various points of view along
  with its various generalizations
  \cite{BerLei,Yil,Buc,JanRobWin,Ell,br73,Wym,Arm,SusZha}.

  An obvious generalization of models with a single scalar field is to
  invoke a set of such fields, $\Phi = (\Phi^1,...,\Phi^n)$. Such
  {\it multi-field models\/} have been applied in inflationary Universe
  theory (see, e.g., \cite{lidlyt00} and references therein). A more
  significant generalization is represented by nonlinear sigma models
  (NSM) in which the fields of the multiplet take part in a geometric
  interaction by forming an inner space (the so-called {\it {\rm target
  space}\/}) with a Riemannian metric $h_{ab}(\Phi)$ \cite{perelomov87}.
  The action of the theory then must be invariant under coordinate
  transformations both in our space-time and in the target space.

  The terms ``chiral model'' and ``chiral fields'' are used to stress 
  that interaction is inserted in a purely geometric way, unlike 
  scalar fields in multi-field models. In the latter, interaction is 
  inserted by simply adding the interaction Lagrangian to the one of
  free fields \cite{perelomov87}. In \cite{perelomov87}, the term 
  ``chiral model'' is used as an equivalent to NSM. A historical comment
  on the term "chiral" can be found in \cite{ch97m}.

  Chiral NSM have been introduced by Schwinger \cite{schwinger57} and Skyrme
  \cite{skyrme58}. Gell-Mann and Levy \cite{gellev60} pointed out how to
  realize the chiral symmetry and partial conservation of the axial vector
  current. A two-dimensional version of the model was studied because of
  its analogy in many respects to non-Abelian gauge theories. The main
  results of these studies can be found in the review \cite{perelomov87}.

  A bridge to four-dimensional NSM was built only with inclusion of
  a coupling to gravity in \cite{aff79}. NSM as a source of gravity were
  also considered by G. Ivanov \cite{ivanov83tmf} (see also \cite{ch83iv}).

  Applications of chiral NSM in inflationary cosmology have been proposed in
  \cite{ch95gc,ch97gc}. Chiral NSM with self-interaction potentials, leading
  to the so-called {\it chiral cosmological models\/}, contain
  self-interacting scalar field theories as well as simple multi-field
  theories as special cases. One can also mention the sigma-model approach to
  a broad class of problems of multidimensional gravity, partly related to
  string and M-theories and leading to many exact solutions, though dealing
  with constant target space metrics --- see \cite{iv-mel} and references
  therein.

  The aim of the present work is to explore NSM as a source of gravity
  in static, spherically symmetric configurations in general relativity.

\section{Field equations}

\subsection{General formalism}

  The action for a self-gravitating chiral model based on a nonlinear sigma
  mode with $n$ scalar fields $\Phi^a$, minimally coupled to gravity, 
  and an interaction potential $V(\Phi)$ has the form
\bearr                         \label{action}
    S= \int d^4x \sqrt{-g} \left\{ R-
    g^{\mu\nu} h_{ab}\Phi^a_{,\mu} \Phi^b_{,\nu}  - 2V(\Phi)\right\},
\ear
  where $g_{\mu\nu}(x)$ is the spacetime metric,\footnote
    {We choose the metric signature ($-,+,+,+$), the units
     $c = 8\pi G =1$, and the sign of $T\mN$ such that $T^0_0$ is the
     energy density.}
  $R$ is the scalar curvature, $h_{ab}(\Phi)$ is a
  metric in the target space, Latin indices run from 1 to $n$, and
  $\Phi = (\Phi^{1},...,\Phi^{n})$; commas and semicolons in the indices
  stand for partial ($\d/\d x^\mu$) and covariant ($\nabla_\mu$) derivatives,
  respectively.

  Varying the action \Ref{action} with respect to the metric
  $g_{\mu\nu}$ gives the Einstein equations
\beq                                \label{eineq}
    R_{\mu\nu} -\half g_{\mu\nu} R = - T_{\mu\nu},
\eeq
  with the stress-energy tensor (SET) of the chiral fields
\bearr\label{ch-em}
    T_{\mu\nu}=h_{ab}\Phi^a_{,\mu}\Phi^b_{,\nu}- g_{\mu\nu}\left[\half
    g^{\alpha\beta}h_{ab}\Phi^a_{,\alpha}\Phi^b_{,\beta}
            +V(\Phi)\right].
\ear
  \eqs \Ref{eineq} can be easily transformed to
\beq                                \label{ein}
    R_{\mu\nu} =h_{ab}\Phi^a_{,\mu}\Phi^b_{,\nu}+ g_{\mu\nu}V(\Phi).
\eeq
  Varying the action (\ref{action}) with respect to the chiral fields
  $\Phi^c$ gives the chiral field equations of motion:
\bearr                                         \label{cfe}
    h_{ab}\Phi^{b;\,\mu}_{\ ;\,\mu}+ \left[\frac{\d
    h_{ab}}{\d\Phi^c} -\frac12 \frac{\d
    h_{bc}}{\d\Phi^a} \right]\Phi^{b;\,\mu}\Phi^c_{;\,\mu}
        -\frac{\d V}{\d\Phi^a}=0.
\ear
  As usual, one should check whether the SET obeys the usual energy
  conditions. In particular, the null energy condition (NEC) reads
  $T_{\mu\nu}k^\mu k^\nu\ge 0$, where $k^\mu$ is an arbitrary null vector.
  For chiral fields with the stress-energy tensor \Ref{ch-em}, the NEC yields
\beq\label{NEC}
    \Xi := h_{ab}\Phi^a_{,\mu} k^\mu \Phi^b_{,\nu} k^\nu \ge 0.
\eeq
  This means that $\Xi$ should be a positive-definite quadratic form with
  respect to the vectors $\zeta^a = \Phi^a_{,\mu} k^\mu$ in the target space.
  Evidently, the NEC is violated if $\Xi < 0$.

\subsection{Two-component sigma model}

  A simple example of a nonlinear sigma model is that consisting
  of two scalar field components, i.e., $\Phi=(\phi,\psi)$. A general
  quadratic form in the target space now is
\beq                              \label{quadform}
  g^{\mu\nu}[h_{11}\phi_{,\mu}\phi_{,\nu}+2h_{12}\phi_{,\mu}\psi_{,\nu}
    +h_{22}\psi_{,\mu}\psi_{,\nu}],
\eeq
  where $h_{ab} = h_{ab}(\phi,\psi)$. After appropriate transformations in
  the target space:  $\phi = \phi(\tilde\phi,\tilde\psi)$, $\psi =
  \psi(\tilde\phi,\tilde\psi)$, the quadratic form \Ref{quadform} can, in
  principle, be reduced to its canonical form
\beq
    g^{\mu\nu}[\phi_{,\mu}\phi_{,\nu}+h\,\psi_{,\mu}\psi_{,\nu}],
\eeq
  where $h$ is, generally speaking, a function of $\phi$ and $\psi$,
  i.e. $h=h(\phi,\psi)$. The especially simple case, which corresponds to
  rotational symmetry in the target space, is $h=h(\phi)$. In this case the
  action \Ref{action} reduces to
\bearr                            \label{action2}
    S= \int d^4x \sqrt{-g} \Bigl\{R-
    g^{\mu\nu}[\phi_{,\mu}\phi_{,\nu}+h(\phi)\psi_{,\mu}\psi_{,\nu}]
\nnn\inch
        -2V(\phi,\psi)\Bigr\}.
\ear
  Now the chiral field $\phi$ can be interpreted as an ordinary scalar field
  and $\psi$ as a scalar field coupled with $\phi$ via the kinetic coupling
  function $h(\phi)$.

\section{Static, spherically symmetric solutions}

\subsection {Equations and geometry}

  In this section we will consider static, spherically symmetric solutions of
  the theory \Ref{action} without a potential, i.e., assuming $V(\Phi)
  \equiv 0$. For a general static spherically symmetric configuration,
  $\Phi^a = \Phi^a(u)$, where $u$ is an arbitrary radial coordinate,
  and the spacetime metric can be written as
\beq                                                           \label{ds}
     ds^2 = -\e^{2\gamma(u)}dt^2 + \e^{2\alpha(u)}du^2
                         + \e^{2\beta(u)} d\Omega^2.
\eeq
  where $d\Omega^2 = (d\theta^2 + \sin^2\theta d\varphi^2)$ is the linear
  element on a unit sphere.

  The SET of the scalar fields has the form
\beq
    T\mN = h_{ab} \Phi^a{}' \Phi^b{}' \diag(1, -1, 1, 1),    \label{SET}
\eeq
  i.e., has the same structure as for a single massless scalar field
  (the prime denotes $d/du$). Therefore the metric has the same form as in
  this simple case and should be reduced to the Fisher metric \cite{Fis} if
  the scalar fields behave as a canonical scalar with positive kinetic energy
  and to the metric of the corresponding solution for a phantom scalar, first
  found by Bergmann and Leipnik \cite{BerLei} (it may be called
  ``anti-Fisher'', by analogy with anti-de Sitter), if the scalar fields
  behave in a phantom way. Let us reproduce this solution in the simplest
  joint form, suggested in \cite{br73}.

  Two combinations of the Einstein equations for the metric (\ref{ds}) and
  the SET (\ref{SET}) read $R^0_0 =0$ and $R^0_0 + R^2_2 =0$.
  Choosing the harmonic radial coordinate $u$, such that $\alpha(u) =
  2\beta(u) + \gamma(u)$, we easily solve these equations. Indeed, the first
  of them reads simply $\gamma'' =0$, while the second one is written as
  $\beta'' + \gamma'' = \e^{2(\beta+\gamma)}$. Solving them, we have
\bearr
     \gamma = - mu,                                       \label{s}
\nnn
     \e^{-\beta-\gamma} =  s(k,u) := \vars     {
                    k^{-1}\sinh ku,  \ & k > 0, \\
                                  u, \ & k = 0, \\
                    k^{-1}\sin ku,   \ & k < 0.     }
\ear
  where $k$ and $m$ are integration constants; two more integration constants
  have been suppressed by choosing the zero point of $u$ and the scale along
  the time axis. As a result, the metric has the form \cite{br73}
\beq  \nq                                                      \label{ds1}
     ds^2 = -\e^{-2mu} dt^2 + \frac{\e^{2mu}}{s^2(k,u)}
                    \biggr[\frac{du^2}{s^2(k,u)} + d\Omega^2\biggl].
\eeq

  In addition, with these metric functions, the ${1\choose 1}$ component of
  the Einstein equations (\ref{eineq}) leads to
\beq                                                           \label{int}
     k^2 \sign k = m^2 + h_{ab} \Phi^a{}' \Phi^b{}',
\eeq
  whence it is clear that
\beq                                                             \label{N}
     h_{ab} \Phi^a{}' \Phi^b{}' = N = \const.
\eeq
  The scalar field equations read
\beq                                                          \label{eq-s}
    2 \big(h_{ab}\Phi^b{}'\big)'
            + \frac{\d h_{bc}}{\d\Phi_a}\ \Phi^b{}' \Phi^c{}' =0
\eeq
  and obviously cannot be solved in a general form, but (\ref{N}) is their
  first integral.

  The metric (\ref{ds1}) is defined (without loss of generality) for $u >0$,
  it is flat at spatial infinity $u = 0$, and $m$ has the meaning of the
  Schwarzschild mass in proper units. The metric properties crucially depend
  on the sign of $k$, which in turn depends on $N$ and ultimately on the
  nature of the matrix $h_{ab}$.

  If $h_{ab}$ is positive-definite, we have $N = k^2 -m^2 > 0$ for all
  nontrivial scalar field configurations and obtain the Fisher metric:
  then $k>0$, and the substitution $\e^{-2ku} = 1 - 2k/r$ converts
  (\ref{ds1}) to
\bearr                                                         \label{ds+}
     ds^2 = -(1 - 2k/r)^{a} dt^2 + (1 - 2k/r)^{-a} dr^2
\nnn \inch
                       + (1 - 2k/r)^{1 - a} r^2 d\Omega^2,
\ear
  where $a = m/k = (\sign m)(1 - N/k^2)^{1/2} < 1$.\footnote
    {Note that the metric \Ref{ds+} has been first
     presented by Buchdahl \cite{Buc}.}
  The solution is defined for $r\!>\! 2k$, $r\!\to\!\infty$ is a flat asymptotic,
  and $r = 2k$ is a central singularity. The Schwarzschild metric is restored
  for $N = 0$, $a=1$. In full agreement with the well-known general results
  \cite{MorTho, hoh-vis}, in this case, due to $\Xi >0$, throats (i.e.,
  minima of the spherical radius $\cR(r) = \e^{\beta}$) are absent, and
  wormholes are impossible.

  If $h_{ab}$ is not positive-definite, $N$ can have any sign, Thus, in
  particular, for some nontrivial scalar field configurations we may have
  $N = 0$, hence the Schwarzschild metric. It is an interesting new feature
  which is absent in the case of a single scalar field. More generally, there
  can be now $N < 0$, which corresponds to ``anti-Fisher'' phantom-field
  metrics. Let us briefly describe different cases of these metrics,
  following \cite{cold08} and the book \cite{BR08}. According to the three
  variants of the function $s(k,u)$ at different $k$ in (\ref{s}), the
  solution with $N < 0$ splits into three branches with the following
  properties.

\medskip\noi
{\bf (a)} $k > 0$: the metric has again the form (\ref{ds+}), but now, due to
    $N < 0$, it holds $|a| > 1$. For $m < 0$, that is, $a < -1$, we have,
    just as in the Fisher solution, a repulsive central singularity at $r=2k$.

    The situation is, however, drastically different for $a > 1$. Indeed,
    the spherical radius $\cR$ then has a finite minimum at $r = r_{\rm
    th} = (a+1)k$, corresponding to a throat of the size
\beq  \nq
    \cR (r_{\rm th}) = \cR_{\rm th} = k (a+1)^{(a+1)/2} (a-1)^{(1-a)/2},
\eeq
    and tends to infinity as $r\to 2k$. Moreover, for $a=2,\ 3,\ \ldots$
    the metric exhibits a horizon of order $a$ at $r = 2k$ and admits a
    continuations to smaller $r$. A peculiarity of such horizons is their
    infinite area. Such \asflat\ configurations with horizons of infinite
    size have been named cold \bhs\ since all of them have zero Hawking
    temperature (see \cite{cold08} and references therein for more detail).

    Furthermore, one can verify that all nonzero components
    $R\mn{}^{\rho\sigma}$ of the Riemann tensor behave as $P^{a-2}$ (where $P
    = 1-2k/r$) as $r\to 2k$ and $P\to 0$.  An exception is the value $a = 1$,
    corresponding to $N = 0$, when the Schwarzschild solution is reproduced.
    Hence, at $r = 2k$ the metric has a curvature singularity if $a < 2$
    (except for $a = 1$), there is finite curvature if $a=1$ and $a=2$ and
    zero curvature if $a > 2$.

    For non-integer $a > 2$, the qualitative behavior of the metric as
    $r \to 2k$ is the same as near a horizon of infinite area, but a
    continuation beyond it is impossible due to non-analyticity of the
    function $P^a(r)$ at $r = 2k$. Since geodesics terminate there at a
    finite value of the affine parameter, it is a space-time singularity
    (a {\it singular horizon\/} as it was termed in \cite{cold08}) even
    though the curvature invariants tend there to zero.

\medskip\noi
{\bf (b)} $k = 0,\ N = -m^2$: the metric is defined in the range $u \in \R_+$
    and is rewritten in terms of the coordinate $r = 1/u$ as
    follows:\footnote
      {To our knowledge, this metric was for the first time found
           by Yilmaz \cite{Yil}.}
\bear
       ds^2 = -\e^{-2m/r}dt^2                                \label{ds0}
                      + \e^{2m/r}[dr^2 + r^2 d\Omega^2],
\ear
    As before, $r = \infty$ is a flat infinity, while at the other
    extreme, $r\to 0$, the behavior is different for positive and
    negative mass. Thus, for $m < 0$, $r =0$ is a singular center ($r=0$,
    the curvature invariants are are infinite). On the contrary, for $m > 0$,
    $r \to \infty$ and the curvature tensor tends to zero as $r \to 0$. This
    is again a singular horizon: despite the vanishing curvature, the
    non-analyticity of the metric in terms of $r$ makes its continuation
    impossible.

    The geometries (\ref{ds+}) with $a > 1$ and (\ref{ds0}) with $m > 0$
    can be called \whs, but they are \asflat\ only as $r\to\infty$, whereas
    on the other side of the throat there is a singular or nonsingular
    horizon of infinite area.

\medskip\noi
{\bf (c)} $k < 0,\ -N = m^2+k^2$: the solution describes a \wh\ with two flat
    asymptotics at $u=0$ and $u = \pi/|k|$. The metric has the form
\bearr                                                          \label{ds-}
    ds^2 = -\e^{-2mu} dt^2 + \frac{k^2\e^{2mu}}{\sin^2 (ku)}
        \biggl[ \frac{k^2\,du^2}{\sin^2 (ku)} + d\Omega^2 \biggr]
\nnnv
       = -\e^{-2mu} dt^2 + \e^{2mu} [dr^2 + (k^2 + r^2) d\Omega^2],
\ear
    where $u$ is expressed in terms of the coordinate $r$, defined on the
    whole real axis $\R$, by $|k| u =  \cot^{-1} (r/|k|)$.  If $m > 0$, the
    \wh\ is attractive for ambient test matter at the first asymptotic ($r\to
    \infty$) and repulsive at the second one ($r \to - \infty$), and vice
    versa in case $m < 0$.  For $m = 0$ one obtains the simplest possible
    \wh\ solution, called the Ellis \wh, although Ellis \cite{Ell} actually
    discussed these \wh\ solutions with any $m$.

    The \wh\ throat occurs at $r = m$ and has the size
\beq
    \cR_{\rm th} = (m^2 + k^2)^{1/2}
                 \exp \left(\frac{m}{k} \cot^{-1}\frac{m}{k}\right).
\eeq

\subsection{Scalar field configurations}

  The metrics \Ref{ds+}, \Ref{ds0}, and \Ref{ds-} describe all possible kinds
  of \ssph\ spacetime geometries for the model \Ref{action}. The scalar field
  configuration can be determined from \eqs (\ref{eq-s}) which, as has been
  noted, cannot be solved in a general form. Let us therefore consider some
  specific examples for the action (\ref{action2}) with two fields $\phi$ and
  $\psi$ and the kinetic coupling $h(\phi)$.

  Then, in terms of the harmonic coordinate $u$, \eqs (\ref{eq-s}) take the
  form
\bear
    2\phi'' - \frac{dh}{d\phi} \psi'{}^2 \eql 0,           \label{phi''}
\\
    (h \psi')' \eql 0,                                     \label{psi''}
\ear
  while their integral (\ref{N}) reads
\beq                                                         \label{N1}
    \phi'{}^2 + h(\phi) \psi'{}^2 = N.
\eeq
  It is easy to verify that \eq (\ref{phi''}) follows from (\ref{psi''}) and
  (\ref{N1}). \eq (\ref{psi''}) leads to
\beq
    h \psi' = C = \const,                                   \label{psi'}
\eeq
  and both (\ref{psi'}) and (\ref{N1}) with $\psi'$ substituted from
  (\ref{psi'}) are easily integrated by quadratures in a general form:
  indeed, substituting $\psi'$ from (\ref{psi'}) into (\ref{N1}), one obtains
  an integrable first-order equation with respect to $\phi(u)$. Moreover, it
  is clear that $N >0$ as long as $h(\phi) \geq 0$ for all nontrivial scalar
  field configurations, and we have the Fisher geometry (\ref{ds+}), $|a| <
  1$. In other words, throats and \whs\ can only be obtained with $h < 0$.

  Let us present three examples of $h(\phi)$ for which the quadratures are
  found explicitly.

\medskip\noi
{\bf 1.} $h(\phi) = 1/\phi$. In this case, integrating (\ref{N1}) with
  (\ref{psi'}), we find
\beq                                                         \label{phi1}
    \phi = \frac{N}{C^2} - \frac{C^2}{4} (u-u_0)^2,
\eeq
  and substituting it into (\ref{psi'}), we finally get
\beq                                                         \label{psi1}
    \psi = \psi_0 + \frac{N}{C} u - \frac{C^3}{12} (u-u_0)^3,
\eeq
  where $u_0$ and $\psi_0$ are integration constants. The main observation is
  that the solution is defined for any $N$, hence all geometries described
  above are possible. In particular, in case $N=0$ we have the Schwarzschild
  geometry despite the existence of two nontrivial scalar fields.

\medskip\noi
{\bf 2.} $h(\phi) = \phi^2$. Integration gives
\bear                                                            \label{phi2}
    \phi \eql \sqrt{N}\sqrt{C^2/N^2 + (u-u_0)^2},
\\                                                               \label{psi2}
    \psi \eql \psi_0 + \arctan \biggl[\frac{N}{C}(u-u_0)\biggr].
\ear

\noi
{\bf 3.} $h(\phi) = \sin^2\phi$, which corresponds to SO(3) symmetry
  of the target space. We get in a similar way
\bear \nhq                                                    \label{phi3}
        \phi \eql \sqrt{1 - C^2/N} \sin \left[\sqrt{N}(u-u_0)\right],
\\    \nhq                                                    \label{psi3}
    \psi \eql \psi_0
    + \arctan \left\{\frac{C}{\sqrt{N}} \tan\left[\sqrt{N}(u{-}u_0)
                \right]\right\}.
\ear
  In the second and third cases, $h\geq 0$ and accordingly $N >0$, i.e., only
  the Fisher metric (\ref{ds+}) with $|a| <1$ takes place.

  In all three cases, expressions for $\phi$ and $\psi$ in terms of the
  radial coordinate $r$ used in the metrics (\ref{ds+}), (\ref{ds0}),
  (\ref{ds-}) are easily obtained by substituting the appropriate expressions
  for $u(r)$.

\section{Concluding remarks}

  We have considered static, spherically symmetric solutions of the theory
  \Ref{action} without a potential, i.e., assuming $V(\Phi) \equiv 0$. 
  In this case, the stress-energy tensor of chiral fields has
  the same algebraic structure as for a single massless scalar field, namely,
  $T\mN = h_{ab} \Phi^a{}' \Phi^b{}' \diag(1, -1, 1, 1).$ As a consequence, a
  geometry of static, spherically symmetric spacetime with chiral fields
  turns out to be the same as in a simple model with a single scalar field;
  its properties are determined by the quadratic form $h_{ab}$. All possible
  kinds of these geometries are described by the metrics \Ref{ds+},
  \Ref{ds0}, and \Ref{ds-}.

  Thus, if it is positive-definite, we obtain the Fisher metric (\ref{ds+})
  \cite{Fis} with a naked singularity, originally found for a canonical
  scalar field with positive kinetic energy. In $h_{ab}$ is
  negative-definite, the metric corresponds to the so-called anti-Fisher
  solution for a phantom scalar, first found by Bergmann and Leipnik
  \cite{BerLei}, including singular and nonsingular horizons of infinite
  area, with the metrics (\ref{ds+}) and (\ref{ds0}) as well as traversable
  wormholes with the metric (\ref{ds-}).

  If $h_{ab}$ is neither positive- nor negative-definite, all branches of
  the (anti-)Fisher solution are possible, depending on a particular scalar
  field configuration. A new interesting feature of NSM that appears in
  this case but is absent with a single scalar field is that the family of
  solutions includes the Schwarzschild metric corresponding to some
  nontrivial chiral field configurations, whose SET is in such cases equal to
  zero.

  Some explicit examples of chiral field configurations have been obtained
  for the action (\ref{action2}) with two fields $\phi$ and $\psi$ and the
  kinetic couplings $h(\phi) = \phi^{-1},\ \phi^2,\ \sin^2\phi$.

  For configurations with nonzero potentials $V(\Phi)$, it is hard to find
  exact solutions but some qualitative results are known, which, taken
  together, give rather a clear picture of what can and what cannot be
  expected from static, minimally coupled NSM in general relativity as well
  as some its extensions, including (multi-)scalar-tensor and
  multidimensional theories \cite{vac5}. Let us enumerate some results
  valid for \ssph\ configurations with the action (\ref{action}), the metric
  (\ref{ds}) and $\Phi^a = \Phi^a(u)$.

\begin{enumerate}    \itemsep 1pt
\item
    The no-hair theorem generalizing that of Adler and Pearson \cite{ad-pear}
    to NSM: it claims that if the matrix $h_{ab}(\Phi)$ is positive-definite
    and the potential $V(\Phi) \geq 0$, \asflat\ \bhs\ cannot have nontrivial
    external scalar fields.
\item
    Nonexistence of particlelike solutions (i.e., \asflat\ solutions with a
    regular center) for NSM with positive-definite $h_{ab}(\Phi)$ and
    $V(\Phi)\geq 0$.
\item
    Nonexistence of regular solutions without a centre (wormholes,
    horns, flux tubes) for NSM with positive-definite $h_{ab}(\Phi)$ and
    {\it any\/} potentials $V(\Phi)$.
\item
    The causal structure theorem generalizing that of \cite{vac1}: it asserts
    that, with {\it any\/} $h_{ab}(\Phi)$ and $V(\Phi)$, the list of possible
    types of global causal structures (and the corresponding Carter-Penrose
    diagrams) is the same as in the trivial case $\Phi^a = \const$, namely:
    Minkowski (or AdS), Schwarzschild, de Sitter and Schwarzschild --- de
    Sitter.
\end{enumerate}

  The fourth statement holds irrespective of any assumptions on the spatial
  asymptotics and admits many generalizations. It is therefore the most
  universal property of self-gravitating scalar fields in various theories of
  gravity \cite{vac5}.

\subsection*{Acknowledgments}
    This work was supported in part by the Russian Foundation for
    Basic Research grants No. 08-02-91307, 08-02-00325, 09-02-0677-a.

\end{document}